\documentclass{article}
\def\Ref#1{(\ref{#1})}
\usepackage{amsmath}
\usepackage{amssymb}
\usepackage{cite}
\def\d{{\rm d}}

\begin{document}
\begin{titlepage}
\noindent{\large\textbf{Models solvable through the empty-interval
method}}

\vskip 2 cm

\begin{center}Amir~Aghamohammadi\footnote
{mohamadi@alzahra.ac.ir} \& Mohammad~Khorrami\footnote
{mamwad@mailaps.org} \vskip 5 mm

\textit{     Department of Physics, Alzahra University,
             Tehran 19938-91167, Iran. }

\end{center}
\begin{abstract}
\noindent The most general one dimensional reaction-diffusion
model with nearest-neighbor interactions solvable through the
empty interval method, and without any restriction on the
particle-generation from two adjacent empty sites is studied. It
is shown that turning on the reactions which generate particles
from two adjacent empty sites, results in a gap in the spectrum of
the evolution operator (or equivalently a finite relaxation time).
\end{abstract}
\begin{center} {\textbf{PACS numbers:}} 05.40.-a, 02.50.Ga

\end{center}
\end{titlepage}
\section{Introduction}
Reaction-diffusion systems have been studied using various
methods, including analytical techniques, approximation methods,
and simulation. The proper approximation methods are generally
different in different dimensions, as for example the mean field
techniques, working good for high dimensions, generally do not
give correct results for low dimensional systems. A large fraction
of analytical studies, belong to low-dimensional (specially
one-dimensional) systems, where solving low-dimensional systems
should in principle be easier.
\cite{ScR,ADHR,KPWH,HS1,PCG,HOS1,HOS2,AL,AKK,RK,RK2,AKK2,AM1}.

In this context, the term solvability (or integrability) is used
in different senses. In \cite{AA,RK3,RK4}, integrability means
that the $N$-particle conditional probabilities' S-matrix is
factorized into a product of 2-particle S-matrices. In
\cite{BDb,BDb1,BDb2,BDb3,Mb,HH,AKA,KAA,MB,AAK,AK}, solvability
means closedness of the evolution equation of the empty intervals
(or their generalization). In \cite{GS,AAMS,SAK}, solvability
means that the evolution equation of $n$-point functions contains
only $n$- or less- point functions.

Among the important aspects of reaction-diffusion systems, are the
stationary state of the system (or one of the quantities
describing the system) and the relaxation behavior of the system
towards this configuration. In the thermodynamic limit (when the
size of the system tends to infinity) these behaviors may show
discontinuity in terms of the control parameters of the system. In
\cite{MA1,AM2,MAM,MA2} (for example), such behaviors are studied.

The empty interval method (EIM) has been used to analyze the one
dimensional dynamics of diffusion-limited coalescence
\cite{BDb,BDb1,BDb2,BDb3}. Using this method, the probability that
$n$ consecutive sites are empty has been calculated. This method
has been used to study a reaction-diffusion process with
three-site interactions \cite{HH}. EIM has been also generalized
to study the kinetics of the $q$-state one-dimensional Potts model
in the zero-temperature limit \cite{Mb}.

In this article, we are going to study all the one dimensional
reaction-diffusion models with nearest neighbor interactions which
can be exactly solved by EIM. It is worth noting that ben-Avraham
et al. have studied one-dimensional diffusion-limited processes
through EIM \cite{BDb,BDb1,BDb2,BDb3}. In their study, some of the
reaction rates have been taken infinite, and they have worked out
the models on continuum. For the cases of finite reaction-rates,
some approximate solutions have been obtained.

We study models with finite reaction rates, obtain conditions for
the system to be solvable via EIM, and then solve the equations of
EIM. In \cite{AKA}, general conditions were obtained for a
single-species reaction-diffusion system with nearest neighbor
interactions, to be solvable through the empty-interval method.
Solvability means that evolution equation for $E_n$ (the
probability that $n$ consecutive sites be empty) is closed. It
turned out there, that certain relations between the reaction
rates are needed, so that the system is solvable via EIM. The
evolution equation of $E_n$ is a recursive equation in terms of
$n$, and that this equation is linear. It was shown that if
certain reactions are absent, namely reactions that produce
particles in two adjacent empty sites, the coefficients of the
empty intervals in the evolution equation of the empty intervals
are $n$-independent, which makes them be solved more easily. The
criteria for solvability, and the solution of the empty-interval
equation were generalized to cases of multi-species systems and
multi-site interactions in \cite{KAA,AAK,AK}.

Here we want to study the case dropped from the study in
\cite{AKA}, namely when there are interactions producing particles
from two adjacent empty sites. Doing so, we are considering the
most general one dimensional reaction-diffusion model with
nearest-neighbor interactions which can be solved exactly through
EIM.

The scheme of the paper is as follows. In section 2, the most
general one dimensional reaction-diffusion model with
nearest-neighbor interactions which can be solved exactly through
EIM is introduced. In the same section the evolution equation of
the empty intervals is obtained for a lattice. Then, using a
limiting procedure a similar equation is obtained for the
continuum. In section 3 the stationary solution to this equation
is obtained. In section four the relaxation of the system towards
its stationary state is investigated. Section 5 is devoted to the
concluding remarks.
\section{Models solvable through the empty interval method}
To introduce the notation, let us briefly review the criteria that
a single-species nearest-neighbor-interaction reaction-diffusion
system be solvable through the empty-interval method (EIM).
Consider a one-dimensional lattice. It was shown in \cite{AKA},
that the most general interactions for a single-species model in a
one-dimensional lattice with nearest-neighbor interactions are
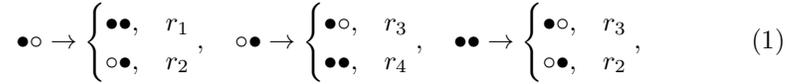
\begin{equation}\label{e1}
  \bullet \circ \to \begin{cases}\bullet \bullet,& r_1\\
                   \circ \bullet,& r_2
                   \end{cases},
\quad  \circ \bullet \to\begin{cases} \bullet \circ,& r_3\\
                               \bullet \bullet,& r_4
                 \end{cases},
\quad  \bullet\bullet   \to \begin{cases}\bullet\circ,& r_3\\
                        \circ \bullet,& r_2
            \end{cases},
\end{equation}
and
\begin{equation}\label{e2}
\circ\circ  \to \hbox{anything},\qquad r,
\end{equation}
in order that the system be solvable through the EIM. Here an
empty (occupied) site is denoted by $\circ$ ($\bullet$), and
$r_i$'s and $r$ are reaction rates. Denoting the probability of
finding n consecutive empty sites by
\begin{equation}\label{e3}
P(\overbrace{\circ\circ\cdots\circ }^n)=:E_n,
\end{equation}
it was then shown that
\begin{align}\label{e4}
  \frac{\d E_n(t)}{\d t}=&(r_2+r_3)(E_{n-1}+E_{n+1}-2E_{n})-
  (r_1+r_4)(E_{n}-E_{n+1})\nonumber\\
  &-(n-1)\,r\,E_n,\qquad n>1,\\ \label{e5}
  \frac{\d E_1(t)}{\d t}=&(r_2+r_3)(1+E_2-2E_1)-
  (r_1+r_4)(E_1-E_2),\\ \label{e6}
  \frac{\d E_{L+1}(t)}{\d t}=&-L\,r\,E_{L+1},
\end{align}
where the length of the lattice has been assumed to be $L+1$. It
is seen that the \Ref{e5} takes a form similar to \Ref{e4},
provided one defines
\begin{equation}\label{e7}
E_0(t):=1.
\end{equation}
In \cite{AKA}, \Ref{e4} to \Ref{e6} were actually obtained for the
case $r=0$. \Ref{e4} to \Ref{e7}, and of course the initial values
of $E_n$'s, are a complete set of equations to obtain $E_n(t)$.
One can absorb the rate $(r_2+r_3)$ in the definition of time, and
rewrite \Ref{e4} and \Ref{e5} as
\begin{align}\label{e8}
  \frac{\d E_n(t)}{\d t}=&(E_{n-1}+E_{n+1}-2E_{n})-
  b\,(E_{n}-E_{n+1})\nonumber\\
  &-(n-1)\,c\,E_n,\qquad 0<n<L+1,
\end{align}
where
\begin{align}\label{e9}
b:=&\frac{r_1+r_4}{r_2+r_3},\nonumber\\
c:=&\frac{r}{r_2+r_3}.
\end{align}
The aim is to solve \Ref{e8} along with \Ref{e6} and \Ref{e7}.

The continuous-space form of the above equations is
\begin{equation}\label{e10}
\frac{\partial E}{\partial t}=\frac{\partial^2 E}{\partial x^2}
+b\,\frac{\partial E}{\partial x}-c\,x\, E, \qquad 0<x<X,
\end{equation}
with the boundary conditions
\begin{align}\label{e11}
E(x=0,t)&=1,\\ \label{e12}
\frac{\partial E(x=X,t)}{\partial
t}&=-c\,X\,E(x=X,t).
\end{align}
The procedure to write these equations in continuous space is to
define
\begin{align}\label{e13}
x&:=n\,\Delta,\nonumber\\
\tilde t&:=\Delta^2\,t,\nonumber\\
\tilde b&:=\frac{b}{\Delta},\nonumber\\
\tilde c&:=\frac{c}{\Delta^3},\nonumber\\
E(x,\tilde t)&:=E_n(t).
\end{align}
One then expands the right-hand sides of \Ref{e7} and \Ref{e8} in
terms of $\Delta$, sends $\Delta$ to zero, and substitutes the
quantities with tilde with the corresponding quantities without
tilde.

Using the new variable $\mathcal{E}$ defined through
\begin{equation}\label{e14}
\mathcal{E}(x,t):=E(x,t)\,\exp\left(\frac{b\,x}{2}\right),
\end{equation}
one can rewrite \Ref{e10} to \Ref{e12} as
\begin{align}\label{e15}
\frac{\partial\mathcal{E}}{\partial t}&=\frac{\partial^2
\mathcal{E}}{\partial x^2}
-\left(\frac{b^2}{4}+c\,x\right)\,\mathcal{E}, \qquad 0<x<X,\\
\label{e16} \mathcal{E}(x=0,t)&=1,\\ \label{e17}
 \frac{\partial \mathcal{E}(x=X,t)}{\partial t}&=-c\,X\,\mathcal{E}(x=X,t).
\end{align}
\section{The stationary solution}
Denote the stationary solution to \Ref{e10} and \Ref{e12} by
$E^{\rm P}$. It is seen that $\mathcal{E}^{\rm P}$ is a linear
combination of the Airy functions, so,
\begin{equation}\label{e18}
E^{\rm P}(x)=\exp\left(-\frac{b\,x}{2}\right)\,\left\{
\alpha\,\mathrm{Ai}\left[c^{-2/3}\,\left(c\,x+\frac{b^2}{4}\right)\right]+
\beta\,\mathrm{Bi}\left[c^{-2/3}\,\left(c\,x+\frac{b^2}{4}\right)\right]\right\},
\end{equation}
where $\alpha$ and $\beta$ are two constants satisfying
\begin{align}\label{e19}
\alpha\,\mathrm{Ai}\left(\frac{c^{-2/3}\,b^2}{4}\right)+
\beta\,\mathrm{Bi}\left(\frac{c^{-2/3}\,b^2}{4}\right)&=1,\nonumber\\
\alpha\,\mathrm{Ai}\left[c^{-2/3}\,\left(c\,X+\frac{b^2}{4}\right)\right]+
\beta\,\mathrm{Bi}\left[c^{-2/3}\,\left(c\,X+\frac{b^2}{4}\right)\right]&=0.
\end{align}
This solution is simplified for $X\to\infty$ (the thermodynamic
limit). As $\mathrm{Bi}(y)$ behaves like the exponential of a
$y^{3/2}$ for $y\to\infty$, it is seen that in the thermodynamic
limit $\beta$ is zero. So,
\begin{equation}\label{e20}
E^{\rm
P}(x)=\frac{1}{\mathrm{Ai}\left(\frac{c^{-2/3}\,b^2}{4}\right)}
\,\exp\left(-\frac{b\,x}{2}\right)\,
\mathrm{Ai}\left[c^{-2/3}\,\left(c\,x+\frac{b^2}{4}\right)\right]
,\qquad X\to\infty.
\end{equation}
It is seen that there is a unique stationary solution.

The above argument is valid for $c\ne 0$. If $c=0$, then
$E(x=X,t)$, and hence $\mathcal{E}(x=X,t)$, is $t$-independent,
and one has
\begin{equation}\label{e21}
E^{\rm P}(x)=\gamma+(1-\gamma)\,\exp(-b\,x),\qquad c=0,
\end{equation}
where $\gamma$ is an arbitrary constant between zero and one. It
is seen that in this case the stationary solution is not unique.
It is also noteworthy that \Ref{e21} is {\em not} the limit of
\Ref{e18} or \Ref{e20} as $c$ tends to zero.
\section{Relaxation towards the stationary solution}
Defining
\begin{equation}\label{e22}
F(x,t):= E(x,t)-E^{\rm P}(x),
\end{equation}
it is seen that the evolution equation for $F$ is the same as that
of $E$, except for the fact that the boundary conditions for $F$
are homogeneous. To calculate $F(x,t)$, one seeks the eigenvalues
and eigenvectors of the evolution operator:
\begin{align}\label{e23}
\epsilon\,f_\epsilon(x)=&\frac{\d^2 f_\epsilon}{\d x^2}
+b\,\frac{\d f_\epsilon}{\d x}-c\,x\, f_\epsilon, \qquad 0<x<X,\\
\label{e24} f_\epsilon(0)=& 0,\\ \label{e25}
\epsilon\,f_\epsilon(X)=&-c\,X\,f_\epsilon(X).
\end{align}
The solution to this is
\begin{align}\label{e26}
f_\epsilon(x)=&\exp\left(-\frac{b\,x}{2}\right)\nonumber\\
&\times\left\{
\alpha\,\mathrm{Ai}\left[c^{-2/3}\,\left(c\,x+\epsilon+\frac{b^2}{4}\right)\right]+
\beta\,\mathrm{Bi}\left[c^{-2/3}\,\left(c\,x+\epsilon+\frac{b^2}{4}\right)\right]\right\},
\end{align}
where where $\alpha$ and $\beta$ are two constants satisfying
\begin{align}\label{e27}
\alpha\,\mathrm{Ai}\left[c^{-2/3}\,\left(\epsilon+\frac{b^2}{4}\right)\right]+
\beta\,\mathrm{Bi}\left[c^{-2/3}\,\left(\epsilon+\frac{b^2}{4}\right)\right]&=0,\nonumber\\
\alpha\,\mathrm{Ai}\left[c^{-2/3}\,\left(c\,X+\epsilon+\frac{b^2}{4}\right)\right]+
\beta\,\mathrm{Bi}\left[c^{-2/3}\,\left(c\,X+\epsilon+\frac{b^2}{4}\right)\right]&=0.
\end{align}
It is seen that the above equations for $\alpha$ and $\beta$ have
nonzero solutions, only for certain discrete values of $\epsilon$.
This means that the spectrum of the evolution operator is
discrete, and there is a gap between zero and the largest nonzero
eigenvalue of the evolution operator. Here too, the solution is
simplified if one considers the thermodynamic limit. In this case,
\begin{equation}\label{e28}
f_\epsilon(x)=\exp\left(-\frac{b\,x}{2}\right)\,
\mathrm{Ai}\left[c^{-2/3}\,\left(c\,x+\epsilon+\frac{b^2}{4}\right)\right],
\qquad X\to\infty,
\end{equation}
where $\epsilon$ is among $\epsilon_n$'s:
\begin{equation}\label{e29}
\epsilon_n:=c^{2/3}\,z_n-\frac{b^2}{4},
\end{equation}
and $z_n$'s are the zeros of the Airy function:
\begin{equation}\label{e30}
\mathrm{Ai}(z_n)=0.
\end{equation}
It is seen that if one tends $c$ to zero and $X$ to infinity, the
spectrum of the evolution operator tends to $(-\infty,-b^2/4)$.
However, if one puts $c=0$ and $X=\infty$, and then solves the
eigenvector equation, another result is obtained. In this case,
\Ref{e23} to \Ref{e25} become
\begin{align}\label{e31}
\epsilon\,f_\epsilon(x)=&\frac{\d^2 f_\epsilon}{\d x^2}
+b\,\frac{\d f_\epsilon}{\d x}, \qquad 0<x,\\
\label{e32} f_\epsilon(0)=& 0,\\ \label{e33}
\lim_{x\to\infty}f_\epsilon(x)=&0.
\end{align}
The solution to these is
\begin{equation}\label{e34}
f_\epsilon(x)=\sinh\left(\sqrt{\epsilon+\frac{b^2}{4}}\,x\right)
\,\exp\left(-\frac{b\,x}{2}\right),
\end{equation}
and the only condition for $\epsilon$ is that $\epsilon$ must be
negative. That is, the spectrum of the evolution operator is
$(-\infty,0)$.
\section{Concluding remarks}
The most general one-dimensional single-species exclusion model
was considered, for which the evolution of the empty-intervals is
closed. The effect of particle creation in two empty adjacent
sites was specially investigated. The stationary solution was
obtained and the relaxation towards this stationary solution was
studied. It was shown that if the rate of particle creation in
adjacent empty sites is nonzero, then the spectrum of the
evolution operator of the empty intervals is discrete. If this
rate is zero and the system is infinite, then the spectrum is
continuous. However, the spectrum depends on whether one finds the
spectrum for the finite system and then tends the size of the
system to infinity, or the spectrum is directly calculated for the
infinite system. In the former case, the largest eigenvalue of the
evolution operator is negative (there is a gap in the spectrum)
and the results of \cite{AKA} are recovered. This means the the
relaxation of the system towards its steady state is exponential,
in other words, the system has a finite relaxation time. In the
latter case, there is no gap in the spectrum and the spectrum
extends to zero. So in this case the relaxation of the system
towards its steady state is not exponential, in other words, the
relaxation time of the system is infinite. This is an example of a
system for which the limit of the spectrum as the size of the
system tends to infinity is different from the spectrum of the
infinite system \cite{ScR}.

\end{document}